\documentclass[11pt]{article}
\usepackage{color}
\usepackage{amsmath,amssymb}
\usepackage{amsfonts}

\newcommand{\pd}{\partial}
\newcommand{\Fc}{\mathcal{F}}
\newcommand{\Rc}{\mathcal{R}}
\newcommand{\Gc}{\mathcal{G}}
\newcommand{\AP}{\alpha^{\prime}}
\newcommand{\diag}{\mathop{\mathrm{diag}}\nolimits}
\newcommand{\const}{\mathrm{const}}

\begin{document}

\title{\textbf{Cosmological perturbations in SFT inspired non-local scalar field models}}

\author{\textbf{Alexey~S.~Koshelev$^{1}$\footnote{alexey.koshelev@vub.ac.be} \
 and \ Sergey~Yu.~Vernov$^{2,3}$\footnote{vernov@ieec.uab.es, svernov@theory.sinp.msu.ru}}\vspace*{3mm} \\
\small $^1$Theoretische Natuurkunde, Vrije Universiteit Brussel, \\
\small The International Solvay Institutes, Pleinlaan 2, B-1050
 Brussels,  Belgium\\
 \small $^2$Instituto de Ciencias del Espacio (ICE/CSIC) and \\
\small  Institut d'Estudis Espacials de Catalunya (IEEC),\\
\small Camp. UAB, Fac. Ci\`encies, T. C5,
 E-08193, Bellaterra, Barcelona, Spain \\
\small $^3$Skobeltsyn Institute of Nuclear Physics, Lomonosov  Moscow State University,\\
\small Leninskie Gory 1, 119991, Moscow, Russia}

\date{ \ }

%

\date{~}

\maketitle

%
\begin{abstract}
We study cosmological perturbations in models with a single
non-local scalar field originating from the string field theory
description of the rolling tachyon dynamics. We construct the equation
for the energy density perturbations of the non-local scalar field and
explicitly prove that for the free field it is identical to a
system of local cosmological perturbation equations in a particular
model with multiple (maybe infinitely many) local free scalar fields.
\end{abstract}


\section{Introduction}

Recently a new class of cosmological models based on the string field
theory (SFT) (for details see reviews~\cite{review-sft}) and the
$p$-adic string theory \cite{padic} emerges and attracts a lot of
attention \cite{IA1}--\cite{abjv}.

Models originating from the
SFT exhibit one general non-standard property, namely, they have terms
with infinite order derivatives, i.e. non-local terms. The higher
derivative terms usually produce the well known Ostrogradski
instability \cite{ostrogradski} (see
also~\cite{AV_NEC})\footnote{Additional phantom solutions, obtained by
the Ostrogradski method in some models can be interpreted as
non-physical ones~\cite{Simon,SW}, which one can treat perturbatively,
i.e. evaluating them using the lower order equations of motion.
In~papers~\cite{Simon,SW} the instability problem is reduced to such choice of effective
theory parameters that the instability turns out to be essential
only at times that are not described in the framework of the effective
theory approximation.}.
The Ostrogradski result is related to higher than two but
finite order derivatives. In the case of infinitely order
derivatives it is possible that instabilities do not appear~\cite{noghosts}.

The SFT inspired cosmological models \cite{IA1} are intensively
considered as models for dark energy (DE). The way to solve the
Friedmann equations with a quadratic non-local scalar field potential,
by reducing them to the Friedmann equations with many non-interacting
free massive local scalar fields, has been proposed
in~\cite{Koshelev07,AJV0711} (see also~\cite{Vernov2010}). The obtained
local fields satisfy the second order linear differential equations.
The masses of all local fields are roots of an algebraic or
transcendental equation, which appears in the non-local model. In the
representation of many scalar fields some of them are normal, whereas
other fields are phantom (ghost) ones. Moreover, local fields
can appear with complex masses squared~\cite{GK,KVexact}.

It is known that the state parameter $w>-1$ can be
described by quintessence models, $w=-1$ corresponds to the
cosmological constant, while $w<-1$ is a characteristic property of
models with a single phantom scalar field. The inequality $w<-1$ means
the violation of the null energy
condition (NEC). Therefore, models with a phantom often
plague by the vacuum quantum instability in the ultraviolet region.
Phantom fields look harmful to the theory and a model with a phantom
scalar field is not acceptable from the general point of view.
 On the
other hand a possibility of existence of the DE with $w<-1$ is  not
excluded experimentally. Indeed, contemporary cosmological
observational data~\cite{data,Komatsu} strongly support that the
present Universe exhibits an accelerated expansion providing thereby an
evidence for a dominating DE component~\cite{review-de}. Recent results of WMAP~\cite{Komatsu} together
with the data on Ia supernovae give the following bounds for the DE
state parameter $ w_{\text{DE}}=-1.02^{+0.14}_{-0.16} $. Note that the
present cosmological observations do not exclude an evolving DE state
parameter $w_{\text{DE}}$.

Due to the presence of phantom excitations non-local models are of
interest for the present cosmology. To construct a stable model with
$w<-1$ one should construct the effective theory with the NEC violation
from the fundamental theory, which is stable and admits quantization.
 It is known that the SFT and the $p$-adic string theory are
UV-complete ones. Thus one can expect that resulting (effective) models
should be free of pathologies.
With the lack of quantum gravity one can give a try to string theory or an
effective theory admitting the UV-completion. This is a hint towards
the SFT inspired cosmological models. { The SFT motivated non-local action of a modified gravity model, which contains only the Ricci scalar and it's derivatives up to arbitrary orders,
 has been proposed in~\cite{BMS}~(see also~\cite{sftnonlocal,sftnonlocalKV,BKMV_perturbation}). The notable feature of this model is an exact non-singular bouncing solution.}
Among cosmological models with
$w<-1$, which have been constructed to be free of instability problem,
we can mention the Lorentz-violating dark energy model~\cite{Rubakov},
the ghost condensation model~\cite{GostCond}--\cite{Creminelli0812} and
the brane-world models~\cite{Brane}.

Cosmological models coming out from the SFT or the $p$-adic string
theory are considered in application to inflation
\cite{Calcagni}--\cite{nongauss} to explain in particular appearance of
non-gaussianities. Such models of inflation generically have the
remarkable property that slow roll inflation can proceed even with an
extremely steep potential~\cite{Cline1}. Furthermore it is shown in
\cite{nongauss} that the parameter of nonlinearity,  which
characterizes the non-gaussianity in the cosmic microwave background,
can be observably large in contrary to the standard inflation scenarios
and observationally distinguishable from Dirac-Born-Infeld inflation
models.

For a more general discussion on the string cosmology and coming out of
string theory theoretical explanations of the
observational data the reader is referred to~\cite{string-cosmo}. Other
models obeying nonlocality and their cosmological consequences are
considered in~\cite{nonlocal}. Note also that linear
differential equations of infinite order were studied in the
mathematical literature long time ago~\cite{davis}--\cite{Carleson}
(see~\cite{noghosts} as a review).

The purpose of this paper is to derive the cosmological perturbation
equations in non-local string field theory inspired dark energy models.
Particular models are inspired by the fermionic SFT and the most well
understood process of tachyon condensation. Namely, starting with a
non-supersymmetric configuration,  the tachyon of the fermionic
string  rolls down towards the non-perturbative
minimum of the tachyon potential. This process represents the non-BPS
brane decay according to Sen's conjecture (see \cite{review-sft} for
details). From the point of view of the SFT the whole picture is not
yet known and only vacuum (space-time constant) solutions were
constructed (see \cite{schnabl} for the bosonic SFT and \cite{AGM} for
the fermionic SFT). The above-mentioned SFT models have at
least two vacua and it is interesting to construct time-dependent
solutions that interpolate between different vacua. These solutions are
called rolling solutions. An effective field theory description
explaining the rolling tachyon in contrary is known and numeric
solutions describing the tachyon dynamics were obtained \cite{AJK}.
This effective field theory description does capture the nonlocality of
the SFT.

Linearizing the latter Lagrangian around the true vacuum one
gets a model which is of main concern in the present paper. In this
paper, we consider a very general form of linearized non-local action
for the scalar field keeping the main ingredient, the function
$\Fc(\Box)$, which in fact produces the nonlocality in question, almost
unrestricted. The only strong restriction we impose is the analyticity
of $\Fc(\Box)$.

The cornerstone of the successive analysis for the linearized model is the possibility to
reformulate it as a model
with many local scalar fields. The key role is played by the
characteristic equation $\Fc(J)=0$. To simplify the succeeding analysis
we assume that all roots of the characteristic equation are simple\footnote{The analysis of scalar perturbations in the case of double roots is presented in~\cite{KVexact}.}. The
local model although being fully equivalent to the non-local model
exhibits unusual properties as it will be demonstrated explicitly in
the paper. For example, coefficients can be complex. This does not
produce problems because local scalar fields are not physical and
normally should not be given an interpretation. However, we stress that
to the best of our knowledge such local cosmological models were not
studied in the literature before\footnote{Similar models in the
Minkowski space have been considered, for instance, in~\cite{PaisU}.}.

The paper is organized as follows. In Section~2 we describe the
non-local non-linear SFT model. In Section~3 we sketch the construction
of background solutions in the linearized model. { In Section 4 we consider  background solutions in
the Friedmann--Robertson--Walker (FRW)  metric.} In Section~5 we derive
gauge invariant perturbation equations and prove  that for a model with
a free non-local scalar field these equations are identical to
equations for perturbations in a local model with many free
non-interacting scalar fields\footnote{For applications of other
multi-field cosmological models and related technical aspects see for
instance \cite{emails}.}. In Section~6 we summarize the obtained
results and propose directions for further investigations.


\section{Model setup}

In an arbitrary metric the four-dimensional action motivated by the
string field theory is as follows~\cite{Koshelev07,AJV0701}:
\begin{equation*}
S_T=\!\int\! d^4x\frac{\sqrt{-g}}{g_o^2}\left(-\frac12\pd_\mu T\pd^\mu
T+\frac{T^2}{2\AP}-\frac1{\AP}{V_{int}(\bar T)}\right).
\end{equation*}
Here $\AP$ is the string length squared, $g_o$ is the open string
coupling constant. The field $\bar T=\Gc(\AP\Box)T$.  The function
$V_{int}(\bar T)$ is an open string tachyon self-interaction, it has no
quadratic in $\bar T$ term.
 We use the signature $(-,+,+,+)$, $g_{\mu\nu}$ is
the metric tensor. The d'Alembertian $\Box$ is applied to scalar
functions and can be written as follows
\begin{equation}
 \Box=D^{\mu}\pd_{\mu}=\frac1{\sqrt{-g}}\partial_{\mu}\sqrt{-g}g^{\mu\nu}\partial_{\nu}
\end{equation}
 and $D_\mu$ being a covariant
derivative. The coordinates are denoted by Greek indices
$\mu,\nu,\dots$ running from 0 to 3. Spatial indexes are $a,b,\dots$
and they run from 1 to 3.

Scalar fields $T$ (primarily associated with the open string tachyon)
and $\bar T$ are dimensionless, while $[\AP]=\text{length}^2$ and
$[g_o]=\text{length}$. Factor $1/\AP$ in front of $V_{int}$ looks unusual and
can be easily removed by a rescaling of fields. For our purposes it is
convenient keeping all the fields dimensionless.

We assume that the function $\Gc(\AP\Box)$ has no zero. The field
redefinition $T_b=\bar T$ yields
\begin{equation}
S_T=\int d^4x\frac{\sqrt{-g}}{g_o^2}\left(-\frac12\pd_\mu
\tilde{T_b}\pd^\mu
\tilde{T_b}+\frac{\tilde{T}_b^2}{2\AP}-\frac1{\AP}{V_{int}(T_b)}\right),
\label{action_model_pre2}
\end{equation}
where $\tilde {T_b}=\Gc(\AP\Box)^{-1}T_b$.

The SFT inspired non-local gravitation models~\cite{IA1} are introduced
as a sum of the SFT action of the tachyon field $T_b$ plus the standard gravity part of the
action. One cannot deduce this form of
the action from SFT, he can just assume the minimal form of gravity
interaction of all string modes. It is convenient to introduce
dimensionless coordinates $\bar{x}_\mu=x_\mu/\sqrt{\AP}$, the dimensionless Newtonian constant
$\bar{G}_N=G_N/\AP=1/(8\pi M_P^2\AP)$, where $M_P$
is the Planck mass, and the dimensionless open string coupling constant
$\bar g_o=g_o/\sqrt{\AP}$. This allows us to rewrite the above action as follows
\begin{equation}
S=\int d^4\bar{x}\sqrt{-g}\left(\frac{\bar{R}}{16\pi
\bar{G}_N}+\frac1{\bar{g}_o^2}\left(-\frac12\pd_\mu \tilde T_b\pd^\mu
\tilde T_b+\frac1{2}\tilde T_b^2-{V_{int}(T_b)}\right)\right), \label{action_model}
\end{equation}
where $\bar{R}$ is the curvature scalar in the coordinates $\bar{x}_\mu$.
Note that $\tilde {T_b}=\Gc(\Box)^{-1}T_b$ in $\bar{x}_\mu$.
In the following formulae, we always use dimensionless coordinates and parameters and omit bars for
simplicity.

 Let us emphasize that the potential of
the field $T_b$ is
\begin{equation*}
V={}-\frac1{2\Gc(0)^2}T_b^2+V_{int}(T_b).
\end{equation*}
{If there exists a constant solution $T_0$, which corresponds to an extremum of the potential $V$, then one can linearize the theory around it.}
Using $T_b=T_0+\tau$, we get
\begin{equation*}
V=V(T_0)-\frac1{2\Gc(0)^2}\tau^2+\frac{V_{int}(T_0)''}2\tau^2+\dots.
\end{equation*}
 Such
linearized cosmological models are of the primary concern in the
present paper. Action (\ref{action_model}) can be rewritten as
\begin{equation}
S=\int d^4x\sqrt{-g}\left(\frac{R}{16\pi
G_N}+\frac1{2g_o^2}\tau\Fc(\Box)\tau-\Lambda\right),
\label{action_model2}
\end{equation}
where
$    \Fc=(\Box+1)\Gc^{-2}-m^2$.
Here $m^2\equiv (T_0)''$ and $\Lambda$ accounts
$\frac{V(T_0)}{g_o^2}$. Nonlocal cosmological models of type
(\ref{action_model2}) with
\begin{equation}
\label{F_SFT} \Fc_{\text{sft}}(\Box)={}-\xi^2 \Box+1-c\:e^{-2\Box},
\end{equation}
were previously considered
in~\cite{AJV0701,AJV0711,MulrunyNunes}\footnote{In~\cite{MulrunyNunes}
for example, it has been shown that solving the non-local equations
using the localization technique (see Section~3 for details) is fully
equivalent to reformulating the problem using the diffusion-like
partial differential equations. One can fix the initial data for the
partial differential equation, using the initial data of the special
local fields. This specifies initial data for a non-linear model, and
these initial data can be (numerically) evolved into the full
non-linear regime using the diffusion-like partial differential
equation.}.

The function $\Fc$ is assumed to be an analytic function of its
argument, such that one can represent it by the convergent series
expansion with real coefficients:
\begin{equation}
\Fc=\sum\limits_{n=0}^{\infty}f_n\Box^n \quad \text{ and } \quad f_n\in\mathbb{R}.
\end{equation}
Equations of motion are
\begin{eqnarray}
G^\mu_\nu\equiv R^\mu_\nu-\frac12R\delta^\mu_\nu &=&\frac{8\pi G_N}{g_o^2}T^\mu_\nu\,,\label{EOJ_g}\\
\Fc(\Box)\tau&=&0\,, \label{EOJ_tau}
\end{eqnarray}
where $G^\mu_\nu$ is the Einstein tensor and $T^\mu_\nu$ is the
energy--mo\-mentum (stress) tensor:
\begin{equation}
\label{Tmunu}
\begin{split}
T^\mu_\nu&=\sum_{n=1}^\infty\frac{f_n}{2}
\sum_{l=0}^{n-1}\left[\pd^\mu\Box^l\tau\pd_\nu\Box^{n-1-l}\tau
+\pd_\nu\Box^l\tau\pd^\mu\Box^{n-1-l}\tau-\right.\\
&{}-\left.\delta^\mu_\nu\left(g^{\rho\sigma}
\pd_\rho\Box^l\tau\pd_\sigma\Box^{n-1-l}\tau+\Box^l\tau\Box^{n-l}\tau\right)\right]-g_o^2\Lambda
\delta^\mu_\nu.
\end{split}
\end{equation}
 It is easy to check that the Bianchi identity
is satisfied on-shell and for $\Fc=f_1\Box+f_0$ the usual
energy--momentum tensor for the massive scalar field is reproduced. Note
that equation (\ref{EOJ_tau}) is an independent equation consistent
with system (\ref{EOJ_g}) due to the Bianchi identity.


\section{Background solutions construction in the linearized model}

While solution construction in the full non-linear model
(\ref{action_model})  is not yet known the classical solutions to
equations (\ref{EOJ_g}) and (\ref{EOJ_tau}) were studied and analyzed
in~\cite{Koshelev07,AJV0701,AJV0711,MulrunyNunes,BarnabyKamran2,Vernov2010}.
Here we just briefly notice the key
points useful for purposes of the present paper.

The well-known Weierstrass theorem implies that  an entire function  $F(z)$, which is
not identically zero, can be presented as the following product. Let $m$ be the order of the zero of $F$ at
$0$, and let $\{z_k\}$ be a list of the non-zero zeroes of $F$
counting multiplicity. Then there exists non-negative integers numbers $p_1,
p_2,...$ and an entire function $Q_0(z)$
 such that $z_k$:
\begin{equation}
\label{WP}
F(z)=z^me^{Q_0(z)}\prod_{k=1}^\infty\left(1-\frac{z}{z_k}\right)e^{Q_{k}(z)},
\end{equation}
where
\begin{equation*}
Q_{k}(z)=\sum_{l=1}^{p_k}
 \frac1l\left(\frac{z}{z_k}\right)^l,
\end{equation*}
 The sequence of  natural numbers $\{p_n\}$ can be chosen in such way that the following relation is satisfied:
\begin{equation*}
\sum\limits_{n=1}^\infty\left|\frac{r}{z_n}\right|^{p_n+1}< \infty \,\,{\mbox{for all }} r>0.
\end{equation*}
Note that $\exp(Q_{k}(z))$ in (\ref{WP}) provide the convergence of the product (see~\cite{Shabat} for details).

The main idea of finding solutions to the equations of motion is to
start with equation (\ref{EOJ_tau}) and to solve it, assuming the
function $\tau_B$ is  a sum of eigenfunctions of the d'Alembert
operator:
\begin{equation}
\tau_B=\!\sum\limits_i\tau_i,\label{tau_sum}
\end{equation}
where
\begin{equation}
\label{equtaui} \Box\tau_i=J_i\tau_i~\mbox{ and }~\Fc(J_i)=0~\mbox{ for
any }~i=1,\dots,N.
\end{equation}
Hereafter we use $N$ (which can be infinite as well) denoting the
number of roots and omit writing explicitly the summation limits over
$i$. Without loss of generality we assume that for any $i_1$ and
$i_2\neq i_1$ condition $J_{i_1}\neq J_{i_2}$ is satisfied. Indeed, if
the sum (\ref{tau_sum}) includes two summands $\tau_{i_{1}}$ and
$\tau_{i_{2}}$, which correspond to one and the same $J_i$, then we can
consider them as one summand $\tau_i\equiv \tau_{i_{1}}+\tau_{i_{2}}$,
which corresponds to $J_i$. We can consider the solution $\tau$  as a
general solution if all roots of $\Fc$ are simple. The analysis is more
complicated in the case of multiple roots~\cite{Vernov2010,KVexact} and we skip this possibility
for simplicity.

The condition
\begin{equation}\label{Betaequ}
\Fc(J)=0,
\end{equation}
which is an algebraic or transcendental equation, is known as the
\textit{characteristic} equation. Note that $J$ is dimensionless. In
this way of solving, all the information is extracted from the roots of
equation~(\ref{Betaequ}), which values do not depend on the metric.
Since equation (\ref{EOJ_tau}) is linear in $\tau$ one can take the
function $\tau_B$, defined by (\ref{tau_sum}), as a solution.

In an arbitrary metric the energy--momentum tensor in (\ref{EOJ_g})
evaluated on such a solution is
\begin{equation}
T_{\mu\nu}=\sum\limits_i\Fc'(J_i)\left\{\pd_\mu\tau_i\pd_\nu\tau_i-\frac{g_{\mu\nu}}{2}\left(g^{\rho\sigma}\pd_\rho\tau_i\pd_\sigma\tau_i+J_i\tau_i^2\right)
\right\}-g_o^2\Lambda g_{\mu\nu},
\label{EOJ_g_onshell}
\end{equation}
where we note the absence of cross terms $\tau_i\tau_j$ for $i\neq j$.
The energy--momentum tensor (\ref{EOJ_g_onshell})  coincides with an energy--momentum tensor of $N$ free
massive scalar fields, which can be obtained from the following local action
\begin{equation}
S_{local}=\int d^4x\sqrt{-g}\left(\frac{ R}{16\pi
G_N}-\Lambda\right)+\sum_i S_i, \label{action_model_local}
\end{equation}
\begin{equation}
S_i={}-\frac{\Fc'(J_i)}{2g_o^2}\int
d^4x\sqrt{-g}\left(g^{\mu\nu}\pd_\mu\tau_i\pd_\nu\tau_i
+J_i\tau_i^2\right).
\end{equation}

Equations (\ref{equtaui}) can be obtained as
the variation of the local action (\ref{action_model_local}) and
therefore, they are not additional conditions on $\tau_i$.
Varying $S_{local}$, one can also obtain  the Einstein equations (\ref{EOJ_g}) with
$T_{\mu\nu}$ given by formula (\ref{EOJ_g_onshell}).
If $\Fc(J)$ has only simple roots, then
action (\ref{action_model_local}) is equivalent to the initial
non-local one (\ref{action_model2}), because any solution to the
equation of motion for the non-local field $\tau$ from action
(\ref{action_model2}) can be written as $\tau=\sum\tau_i$, where
$\tau_i$ are solutions to equations of motion from action (\ref{action_model_local}), and Hamiltonians are
the same.

 If $\Fc(J)$ has simple real roots, then values of its first derivative $\Fc'(J)$
is not equal to zero at points $J_i$, moreover positive
and negative values of $\Fc'(J_i)$ alternate, so we can obtain phantom
fields.

As an example one can take
$\Fc(J)=(J+1)e^J$ with only one root $J=-1$. Physically this statement
is easy to understand: theories with only a single pole in the
propagator describe only one physical degree of freedom and hence the
non-local structure does not spoil the system with new spurious ghost
states\footnote{If $\Fc(J)$ has no zeros at all like for example
$\Fc(J)=e^J$, then the underlying field theory has no physical
excitations at all.}~\cite{noghosts}. At the same time, for $\Fc(J)$
with two or more simple real roots we obtain that the non-local model
with action (\ref{action_model2}) contains ghost-like excitations. Note
that the use of truncated function
\begin{equation}
\label{trunc}
  \hat{\Fc}(\Box) \equiv \sum_{n=0}^{\hat{N}} f_n \Box^n
\end{equation}
instead of $\Fc(\Box)$ as an approximation is not correct, because
$\hat{\Fc}$, which is the $\hat{N}$-th degree polynomial in $\Box$, can
contain spurious zeros which are not present in $\Fc(\Box)$. Hence the
corresponding solution for $\tau$ can contain modes $\tau_i$ which are
not presented in the full theory~(see, for example, \cite{AJV0701}).
The detailed analysis of the initial value problem for such non-local
equations and deeper analysis of their mathematical properties can be
found in~\cite{noghosts}.

\section{The Friedmann--Robertson--Walker  metric}
All the above formulae are valid for an
arbitrary metric. Let us consider as a background metric
the spatially flat FRW metric of the form
\begin{equation}
\label{mFr} ds^2={}-dt^2+a^2(t)\left(dx_1^2+dx_2^2+dx_3^2\right),
\end{equation}
where $a(t)$ is the scale factor, $t$ is the cosmic time. Some
useful quantities in this metric read
\begin{equation*}
\begin{split}
\Gamma_{ab}^0&=Hg_{ab},\qquad\Gamma^a_{b0}=H\delta^a_b,\qquad \Box=-\pd_t^2-3H\pd_t+\frac1{a^2}\pd^b\pd_b,\\
R_{\mu\nu}&=\left(\begin{array}{cc}-3(\dot H+H^2)&0\\0&g_{ab}(\dot
H+3H^2)\end{array}\right), \qquad   R=6\left(\dot H+2H^2\right),
\end{split}
\end{equation*}
where $H=\dot a/a$ and a dot hereafter in this paper denotes a
derivative with respect to the cosmic time~$t$. Background solutions
for $\tau$ are taken to be space-homogeneous as well. The
energy--momentum tensor in (\ref{EOJ_g}) in this metric can be written
in the form of a perfect fluid $T^\mu_\nu=\diag(-\varrho,p,p,p)$, where
\begin{equation}
\begin{split}\varrho&=\frac1{2}\sum_{n=1}^\infty f_n\sum_{l=0}^{n-1}
\left(\pd_t\Box^l\tau\pd_t\Box^{n-1-l}\tau+
\Box^l\tau\Box^{n-l}\tau\right)+g_o^2\Lambda,\\
p&=\frac1{2}\sum_{n=1}^\infty f_n\sum_{l=0}^{n-1}\left(\pd_t\Box^l\tau\pd_t\Box^{n-1-l}\tau-
\Box^l\tau\Box^{n-l}\tau\right)-g_o^2\Lambda.
\end{split}
\label{ep_sol}
\end{equation}

Using the above notations we get equation (\ref{EOJ_g}) in the
following form:
\begin{equation}
\begin{split}
3H^2&={8\pi G}\varrho,\quad
\dot H={}-{4\pi G}(\varrho+p),\\
\end{split}
\label{FrEOM}
\end{equation}
where the constant $G\equiv {\bar G}_N/{\bar g}_o^2=G_N/g_o^2$.
The consequence of (\ref{FrEOM}) is the conservation equation:
\begin{equation}
\dot\varrho+3H(\varrho+p)=0. \label{EQUrho}
\end{equation}
Note that system (\ref{FrEOM}) is a non-local and non-linear system of
equations. At the same time using formulae (\ref{ep_sol}) for the
energy density and pressure  it is possible to generate local systems
out of~(\ref{FrEOM}), corresponding to particular solutions of the
initial non-local system. Sometimes it gives a possibility to find
exact analytic solutions to the initial non-local
system~\cite{AJV0711}.

For $\tau=\tau_B$ formula (\ref{ep_sol}) gives:
\begin{equation}
\varrho=\frac{1}{2}\sum_i\Fc'(J_i)\left(\dot\tau_i^2
+J_i\tau_i^2\right)+g_o^2\Lambda,\qquad
p=\frac{1}{2}\sum_i\Fc'(J_i)\left(\dot\tau_i^2
-J_i\tau_i^2\right)-g_o^2\Lambda. \label{EQUFr1}
\end{equation}
Therefore, we can rewrite system (\ref{FrEOM}) as follows:
\begin{equation}
\label{FrEOMtau}
\begin{split}
3H^2&=4\pi G\sum_i\Fc'(J_i)\left(\dot\tau_i^2 +J_i\tau_i^2\right)+8\pi
G_N\Lambda,\\ \dot H&=-4\pi G
\sum_i\Fc'(J_i)\dot\tau_i^2.\\
\end{split}
\end{equation}
 Here number of local scalar
fields is equal to the number of roots of the characteristic equation
(\ref{Betaequ}). If $\Fc(J)$ has infinite number of roots, the
local action (\ref{action_model_local}) has infinite number of
local scalar fields. To find a particular background solution one is allowed
to consider finite number of roots $J_i$ assuming that only local
fields, which correspond to these roots, are non-trivial.

It may turn out that if $\Fc(\Box)$ has an infinite Taylor series
and a finite number of roots, then the corresponding non-local system
is equivalent in the above sense to the system with a finite number of
local fields.



\section{Cosmological perturbations with single \textit{non}-local scalar field}

\subsection{Perturbing the field equations}

The main problem of the present paper is the derivation of
cosmological perturbation equations in models with a non-local scalar
field. We consider the linearized models because they are much better
understood and, what is even more important, several background
solutions are known.

For the linearized model (\ref{action_model2}), we  consider the
background solution as given by (\ref{tau_sum}).
Perturbing the equation of motion for $\tau$, we get
\begin{equation}
\delta(\Fc\tau)=\sum_{n=0}^\infty f_n\delta(\Box^n\tau)=0.\label{dtau}
\end{equation}

Using (\ref{tau_sum}) and the following relation
\begin{equation}
\label{deltabox1}
\delta(\Box^n\tau)=\Box^n\delta\!\tau+\sum_{m=0}^{n-1}\Box^m(\delta\Box)\Box^{n-1-m}\tau_B,
\end{equation}
one has
\begin{equation}
\label{deltabox2}
\delta(\Box^n\tau)=\Box^n\delta\!\tau+\sum_i\frac{\Box^n-J^n_i}{\Box-J_i}(\delta\Box)\tau_i.
\end{equation}

{So, Eq.~(\ref{dtau}) can be written as follows
\begin{equation}
\delta(\Fc\tau)=\sum_i\Fc(\Box)\delta\!\tau_i+\frac{\Fc(\Box)}{\Box-J_i}(\delta\Box)\tau_i=0\,,
\label{dtauexplicit}
\end{equation}
where we put $\delta\!\tau=\sum\limits_i\delta\!\tau_i$ and use that $\Fc(J_i)=0$ for all $J_i$.
Note that the function $\Fc(\Box)/(\Box-J_i)$ has no pole.}

  It follows from (\ref{EOJ_g_onshell}) that if for
some $J_k$ a background solution {$\tau_B$, given by (\ref{tau_sum}),} contains $\tau_k=0$,
then $\delta\!\tau_k$, contributes only to the second order in the
energy--momentum tensor perturbations. In this paper, we consider perturbations
only to the first order, and, therefore, for all $\tau_k=0$   we can put
$\delta\!\tau_k=0$ without loss of generality. If $\Fc$ has an infinite number
of roots, but we select as a background the function
 $\tau$, which includes only a  finite number of $\tau_k$, then only a finite
number of perturbations   $\delta\!\tau_k$ give contribution to the first
 order perturbation equations, whereas in the second order all perturbations are
important.

\subsection{Perturbations of the non-local scalar field  energy--momentum tensor}

{
Let us consider the perturbations of $T^{\mu}_\nu$ in the neighbourhood of the background solution $\tau_B$, which depends only on time.
Substituting into expression~(\ref{Tmunu})
\begin{equation*}
\tau=\tau_B(t)+\delta\tau(t,x^a),
\end{equation*}
 we obtain\footnote{We choose the spatial flat FRW metric as the background and $g^{00}$ means the background value of this component of the metric tensor. $\delta  T^a_a$ is the perturbation of the (a,a) component of $T^\mu_\nu$, no summation.}:
\begin{eqnarray}
&\delta  T^0_0&=\frac1{2}\sum_{n=1}^\infty
f_n\sum_{l=0}^{n-1}\left(g^{00}\left[\pd_0\delta(\Box^l\tau)\pd_0\Box^{n-1-l}\tau_B
+\pd_0\Box^l\tau_B\pd_0\delta(\Box^{n-1-l}\tau)\right]-{}
\right.\label{dT00}\\
&&\left.{}-2\delta g^{00}\pd_0\Box^l\tau_B\pd_0\Box^{n-1-l}\tau_B+\delta(\Box^l\tau)\Box^{n-l}\tau_B
+\Box^l\tau_B\delta(\Box^{n-l}\tau)\right)\nonumber,\\
&\delta  T^a_a&=\frac{1}{2}\sum_{n=1}^\infty
f_n\sum_{l=0}^{n-1}\left(-g^{00}\left[\pd_0\delta(\Box^l\tau)\pd_0\Box^{n-1-l}\tau_B
+\pd_0\Box^l\tau_B\pd_0\delta(\Box^{n-1-l}\tau)\right]-
{}\right.\label{dTaa}\\
&&\left.{}-2\delta g^{00}\pd_0\Box^l\tau_B\pd_0\Box^{n-1-l}\tau_B-\delta(\Box^l\tau)\Box^{n-l}\tau_B-\Box^l\tau_B
\delta(\Box^{n-l}\tau)\right)\nonumber,\\
&\delta T^0_a&=\sum_{n=1}^\infty f_n\sum_{l=0}^{n-1}
\pd^0\Box^l\tau_B\pd_a\delta(\Box^{n-1-l}\tau)=\pd_a\left[\sum_{n=1}^\infty f_n\sum_{l=0}^{n-1}
g^{00}\pd_0\Box^l\tau_B\delta(\Box^{n-1-l}\tau)\right]\label{dT0a},\\
&\delta  T^a_b&=0, \qquad a\neq b. \label{dTab}
\end{eqnarray}
Note that we do not use the specific form of $\tau_B$, given by (\ref{tau_sum}), in these formulae.   }

\subsection{Metric perturbations}
{
Ten independent metric perturbations can be divided into four scalar, four vector
and two tensor perturbations, according to their transformation
properties with respect to three-space coordinate transformations on
the constant-time hypersurface~\cite{Lifshitz1946} (see also~\cite{lk}).
Different types of perturbations do
not mix at the first order~\cite{Bardeen,Mukhanov,hwangnoh}.

Scalar metric perturbations are given by four arbitrary scalar
functions $\alpha(\eta,x^a)$, $\beta(\eta,x^a)$, $\varphi(\eta,x^a)$, $\gamma(\eta,x^a)$ in the following way
\begin{equation}
ds^2=a(\eta)^2\left(-(1+2\alpha)d\eta^2-2\pd_a \beta d\eta
dx^a+((1+2\varphi)\delta^a_b+2\pd_a\pd_b\gamma)dx^adx^b\right),
\label{deltametric}
\end{equation}
where $\eta$ is the conformal time related to the cosmic one as
$a(\eta)d\eta=dt$.
There exist two independent gauge-invariant variables (the
Bardeen potentials), which fully determine the scalar perturbations of
the metric tensor~\cite{Bardeen}:
\begin{equation}
\Phi=\alpha-\dot \chi,\qquad\Psi=H\chi-\varphi, \label{GIvarsBardeen}
\end{equation}
where $\chi\equiv a\beta+a^2\dot\gamma$.
The gauge invariant variables $\Phi$ and $\Psi$ have a very simple physical interpretation: they are  amplitudes
of the metric perturbations in the longitudinal
(conformal-Newtonian) gauge, defined by conditions $\beta=\gamma=0$.

The perturbation functions are as usually Fourier
transformed with respect to the spatial coordinates $x^a$ having
thereby the following form:
\begin{equation*}
\Phi(\eta,x^a)=\Phi(\eta,k)e^{ik_ax^a},\qquad \Psi(\eta,x^a)=\Psi(\eta,k)e^{ik_ax^a}.
\end{equation*}
 Here $k=\sqrt{k_ak^a}$
is the comoving wavenumber. Appearance of just simple partial
derivatives in (\ref{deltametric}) and exponents $e^{ik_ax^a}$ reflects
the fact that the spatial curvature is zero.
 Although the
metric perturbations are defined in the conformal time frame in the
sequel the cosmic time $t$ will be used as the function argument and
all the equations will be formulated with $t$ as the evolution
parameter.

The non-local scalar field
energy--momentum tensor can be rewritten in fluid like
quantities. The energy--momentum tensor of  any perfect fluid can be
parameterized in the conformal time frame as follows (we include the background and scalar perturbations in the formula):
\begin{equation}
T^0_0={}-(\varrho+\delta\!\varrho),\quad
T^0_a={}-\frac1k(\varrho+p)\pd_a v^s,\quad T^a_b=(p+\delta\!
p)\delta^a_b+\left(\frac{\pd^a\pd_b}{k^2}+\frac{\delta^a_b}3\right)\pi^s\,,
\label{deltastresstext}
\end{equation}
where $v^s$ is the velocity or the flux related variable and $\pi^s$ is
the anisotropic stress. The perturbation functions in $T^\mu_\nu$ are
as follows:
$\delta\!\rho(\eta,x)=\delta\!\rho(\eta,k)e^{ik_ax^a}$ and
similar for $\delta\! p$, $v^s$ and $\pi^s$.

From action (\ref{action_model2}) we get to the background order the
energy
density $\varrho$ and the pressure  $p$ given by (\ref{ep_sol}).
From (\ref{dTab}) one gets that $\pi^s=0$. Formulae (\ref{dT00}), (\ref{dTaa}), and (\ref{dT0a})  give the explicit form of
$\delta\!\varrho$, $\delta\!p$, and $v^s$ correspondingly. Note that perturbations $\delta T^\mu_\nu$ are described by scalar functions only, therefore,
 the considering non-local scalar field does not give a contribution in the tensor and vector perturbations.}

\subsection{The scalar perturbation equations}

The following notations will be used in the sequel:
\begin{equation*}
w\equiv \frac{p}{\varrho},\qquad c_s^2\equiv\frac{\dot p}{\dot\varrho},\qquad
e\equiv\delta\! p-c_s^2\delta\!\varrho,
\end{equation*}
where $w$ is the equation of state parameter, $c_s^2$ is the speed of
sound\footnote{This definition of the speed of sound \cite{hwangnoh}
conforms with the canonical one (derivative of the pressure density
w.r.t. the energy density at constant entropy) for perfect fluids but
gives different result for scalar fields. It is convenient, however,
keeping this notation $c_s^2$ for scalar fields as well while it is not
really the physical ``speed of sound''.}. Constant $w$ obviously
results in $c_s^2=w$ and $e=0$. Non-zero $e$ describes entropic
perturbations.

Following the lines of Bardeen's paper~\cite{Bardeen} we define gauge
invariant quantities\footnote{In the longitudinal (conformal-Newtonian)
gauge $\chi=0$, hence, $v_\chi=v^s$.}
\begin{equation}
v_\chi=v^s-\frac ka\chi,\qquad\varepsilon=\frac{\delta\!\varrho}{\varrho}+3(1+w)H\frac{a}{k} v.
\label{GIvars}
\end{equation}

Starting with the Einstein equations (\ref{EOJ_g}),
one yields the following equations for scalar perturbations in
the case $\pi^s=0$
\begin{eqnarray}
&&\Psi=\Phi, \label{deltaGI1}
\\
&&\Psi={}-4\pi G\varrho\frac{a^2}{k^2}\varepsilon\,, \label{deltaGI2}
\\
&&\dot v_\chi+Hv_\chi=\frac k{a(1+w)}\left[\frac
e\varrho+c_s^2\varepsilon+\Phi(1+w)\right], \label{deltaGIconsa}
\\
&&\dot \varepsilon-3Hw\varepsilon+\frac ka(1+w)v_\chi=0.
\label{deltaGIcons0}
\end{eqnarray}
Now one can express $v_\chi$ from the latter equation, express $\Phi$
through $\varepsilon$, using (\ref{deltaGI1}) and (\ref{deltaGI2}), and
substitute all of this into (\ref{deltaGIconsa}). This results in a
single second order differential equation:
\begin{equation}
\ddot\varepsilon+\dot\varepsilon
H(2+3c_s^2-6w)+\varepsilon\left(\dot H(1-3w)
-15H^2w+9H^2c_s^2+\frac{k^2}{a^2}c_s^2\right)={}-\frac{k^2e}{a^2\varrho}.
\label{deltaGIeps0}
\end{equation}

Note that equation (\ref{deltaGIeps0}) is valid for any perfect fluid
with $\pi^s=0$ and is well known in the theory of cosmological
perturbations\footnote{One can check this equation against equation (4.9) in
the Bardeen paper~\cite{Bardeen}. Our's and Bar\-deen's formulae are in
a perfect agreement with each other. To do this comparison one has to
account that dot in Bardeen's paper denotes a derivative with respect
to the conformal time, our $e$ is equal to $P_0\eta$ in \cite{Bardeen}
and our $\pi^s$ is equal to $P_0\pi_T^{(0)}$ in~\cite{Bardeen}.}.

 It is suitable to rewrite $e$ in the following form:
\begin{equation}
e=(1-c_s^2)\varrho\varepsilon+\Delta, \label{eDelta}
\end{equation}
 where
\begin{equation}
\Delta=\delta\! p-\delta\!\varrho+(1-c_s^2)\frac ak
\dot\varrho v^s
\end{equation}
and get equation (\ref{deltaGIeps0}) in the following form
\begin{equation}
\ddot\varepsilon+\dot\varepsilon
H(2+3c_s^2-6w)+\varepsilon\left(\dot
H(1-3w)-15H^2w+9H^2c_s^2+\frac{k^2}{a^2}\right)={}-\frac{k^2}{a^2}\frac{\Delta}{\varrho}.
\label{deltaGIeps01nlscalar}
\end{equation}

It is easy to check that $\Delta=0$ in a model with a single local
scalar field. For the non-local scalar field we have
\begin{equation*}
\Delta= -\sum_{n=1}^\infty
f_n\sum_{l=0}^{n-1}\left(\delta(\Box^l\tau)\Box^{n-l}\tau_B
+\Box^l\tau_B\delta(\Box^{n-l}\tau)\right)+(1-c_s^2)\frac { \dot\varrho}{\varrho+p} \sum_{n=1}^\infty
f_n\sum_{l=0}^{n-1} \pd_t\Box^l\tau_B\delta(\Box^{n-1-l}\tau).
\end{equation*}
This does not seem to be equal to zero. Moreover, one should not expect
any significant simplification just because the system with one
non-local scalar field is equivalent to the background order to a
system with many local scalar fields. In a general situation one non-local scalar field
 can correspond to infinitely many local scalar fields.
 This situation is rather complicated. It does not seem
that a description with only one but a non-local scalar field may bring
to us more beautiful equations for perturbations.

Using\footnote{{Note that not $\delta\varrho(\tau_B)$ and $\delta
p(\tau_B)$ separately, only their difference is important for perturbation equations.}}
\begin{equation}
  \delta\varrho(\tau_B)- \delta
p(\tau_B)=2\sum\limits_{i}J_i\Fc'(J_i)\tau_i\delta\!\tau_i,
\qquad  v^s(\tau_B)=\frac k
{a(\varrho+p)}\sum\limits_{i}
J_i\Fc'(J_i)\dot\tau_i\delta\!\tau_i,
 \label{vs}
 \end{equation}
one can get the following expression for $\Delta$
\begin{equation}
\Delta={}-\frac{2}{\varrho+p}\sum_{m}\sum_{l}\Fc'(J_m)\Fc'(J_l)J_m\tau_m\dot\tau_m\dot\tau_l^2\zeta_{ml},
\label{explicitDelta}
\end{equation}
where $\zeta_{ij}=\frac{\delta\tau_i}{\dot\tau_i}-\frac{\delta\tau_j}{\dot\tau_j}$.
In more symmetric form (\ref{explicitDelta}) is as follows:
\begin{equation*}
\Delta=\frac{2}{\varrho+p}\sum_{l}\!\sum_{m>l}\Fc'(J_m)\Fc'(J_l)\dot\tau_m\dot\tau_l\zeta_{ml}
(J_l\tau_l\dot\tau_m-J_m\tau_m\dot\tau_l).\!
\end{equation*}

For example, for $N=2$ we get
\begin{equation*}
\Delta_2=\frac{2}{\varrho+p}\Fc'(J_1)\Fc'(J_2)\dot\tau_2\dot\tau_1\zeta_{21}(J_1\tau_1\dot\tau_2-J_2\tau_2\dot\tau_1).
\end{equation*}

In the case $N>2$ it is worth noting that \textit{despite of the fact that $\frac12N(N-1)$
\textit{nontrivial} functions $\zeta_{ij}$ can be constructed} only $N-1$  functions $\zeta_{ij}$ are
truly independent thanks to the property
\begin{equation}
\zeta_{im}=\zeta_{ij}+\zeta_{jm}.
\end{equation}
 For example, we can consider as independent functions the
functions $\zeta_{1j}$, where $1<j\leqslant N$.

Each local scalar field satisfies the following equation
\begin{equation}
\label{equtau}
\Box\tau_i=J_i\tau_i.
\end{equation}
One can perturb the latter equation and get
\begin{equation*}
\ddot{\delta\!\tau\!}_i+3H\dot{\delta\!\tau\!}_i+\frac{k^2}{a^2}\delta\!\tau_i
+J_i\delta\!\tau_i=2\Phi\ddot\tau_i+(\kappa+3H\Phi+\dot\Phi)\dot\tau_i
\end{equation*}
where $\kappa=3(\dot\Psi+H\Phi)$. Subtracting
two of such equations for indexes $i$ and $j$ one may get
\begin{equation}
\begin{split}
&\ddot{\zeta}_{ij}+\left(3H+\frac{\ddot\tau_{i}}{\dot\tau_i}+
\frac{\ddot\tau_{j}}{\dot\tau_j}\right)\dot\zeta_{ij}+\left(-3\dot H+\frac{k^2}{a^2}\right)\zeta_{ij}=\\
=&\left[\frac{J_i\tau_{i}}{\dot\tau_i}-\frac{J_j\tau_{j}}{\dot\tau_j}\right]
\left(\sum_m\frac{\Fc'(J_m)\dot\tau_m^2}{\varrho+p}(\dot\zeta_{im}+
\dot\zeta_{jm})+\frac{2\varepsilon}{1+w}\right).
\end{split}
\label{deltaijeps}
\end{equation}

These equations together with
\begin{equation}
\begin{split}
&\ddot\varepsilon+\dot\varepsilon
H(2+3c_s^2-6w)+\varepsilon\left(\dot
H(1-3w)-15H^2w+9H^2c_s^2+\frac{k^2}{a^2}\right)={}\\
{}&=\frac{2k^2}{a^2\varrho(\varrho+p)}\sum_{m,l}
\Fc'(J_m)\Fc'(J_l)J_m\tau_m\dot\tau_m\dot\tau_l^2\zeta_{ml}
\end{split}
\label{deltaepsij}
\end{equation}
allow to find $\varepsilon(t)$ and, using equation (\ref{deltaGI2}), the Bardeen potential $\Psi$.

Comoving curvature perturbations can be expressed as
\begin{equation}
  \Rc=\Psi-\frac{H}{\dot H}(\dot\Psi+H\Psi),\label{isocurv}
\end{equation}
entropy perturbations can be found as
\begin{equation}
\frac e{\varrho}=\varepsilon-(1+c_s^2)\frac{a^2}{k^2}\Delta
  \label{entropy}
\end{equation}
Both quantities are gauge invariant and play crucial role in computing
various spectral indexes.

Juxtaposing equations (\ref{deltaijeps}) and equation (\ref{deltaepsij}) with equations from Appendix~1 we see\footnote{All the relevant notations and standard equations for perturbations in
cosmological models with several local scalar fields are
summarized
 in Appendix~1. Note that equations derived
in Appendix are valid only for $k\neq0$ and the zero mode $k=0$
{should be considered separately} (we put the consideration of this case in Appendix~2).}
that perturbations become equivalent in the model with one non-local
scalar field and in the model with many local scalar fields.

If the function $\Fc$ has infinite number of roots and the background solution $\tau_B$ includes
infinite number of $\tau_i$, then system (\ref{deltaijeps})--(\ref{deltaepsij}) consists of infinite number of equations.
We also point out, that in the analysis of the second order perturbation equations, perturbations, which correspond to  $\tau_i=0$
should be taken in the account as well, so, a non-local model with $\Fc$, which has infinite number of roots, is not equivalent to a local model with
a finite number of scalar fields.


Model with a scalar field and a phantom scalar field and quadratic potential has been studied in~\cite{Brandenberger08}.
Perturbations in the neighborhood of the bounce (rolling) solution has been analysed.  Note that the local action,
considering in~\cite{Brandenberger08} can be obtained from the non-local action~(\ref{action_model2}) if $\Fc$ has both zero and positive roots.


\section{Summary and outlook}

The main result of this paper is the construction of the perturbation
equations in non-local models and the explicit proof that the
cosmological model with one free non-local scalar field is equivalent
to a cosmological model with many free local scalar fields not only to
the background but also to the linear perturbation order. The non-local
model is described by action (\ref{action_model2}) and the
corresponding local model is described by action
(\ref{action_model_local}). The latter local model contains $N$ scalar
fields where $N$ is the number of roots of the characteristic equation
$\Fc(J)=0$. Masses squared of these local fields are exactly roots of
the characteristic equation. Perturbation equations for this local
model are (\ref{deltaijeps}) and (\ref{deltaepsij}) where only $N-1$
functions $\zeta_{1j}$ are independent. The obtained perturbation equations
are valid in the case, when $\Fc(J)$ has an infinite number of roots as well.
In this case we get an infinite system of the second order differential equations.
If we choose the background solution $\tau_B$ as a finite sum of $\tau_i$, then only
a finite number of the obtained perturbations equations are nontrivial ones, therefore,
we get a finite system of local equations. Note that such localization can be obtained only for
the first order perturbation equations.

In fact most interesting
exactly known cosmological solutions incorporate the cosmological
constant as an additional
ingredient. We refer reader to \cite{IA1,Koshelev07} on the
discussion on how such a constant can be generated during the tachyon
evolution. In the case of only
one root of the characteristic equation and consequently only one
scalar field in the local model one finds that the R.H.S. of
(\ref{deltaepsij}) is equal to zero and one is left with equations for
perturbations as they are in a local system with a single scalar field.
Moreover, if some fields in the local model are taken trivial in the
background related to them perturbations turn out to be trivial as
well.

Thus the problem of cosmological perturbations in a non-local scalar
field theory is reduced to the problem of cosmological perturbations in
a local theory with many degrees of freedom. In particular,
perturbations in a quintom model very close to our setup with a phantom
field without potential and an ordinary scalar field with quadratic
potential were studied in~\cite{Brandenberger08}. Perturbations in
models with many scalar fields were studied in literature considering
various cosmological scenarios \cite{manyperturbations}. The
characteristic feature of the present setup is that all the local
fields in fact are not physical and play a role of auxiliary functions
introduced for the reduction of the complicated non-local problem to a
known one. Partly because of this artificial origin, the local
counterpart is not always the one already studied. As it was noted in
\cite{Koshelev07,AJV0701} for a very wide class of SFT inspired
functions $\Fc(J)$ infinite number of complex roots $J_i$ may appear.
Looking strange they do not produce a problem for the model since they
are not physical quantities. The corresponding $\tau_i$ also become
complex but it is a matter of choice of integration constants to make
the physical quantities, $\tau$, $\varrho_\tau$ and their perturbations
real.

In the next paper~\cite{KVexact}, we consider one indicative example with one
pair of complex conjugate roots. We demonstrate numerically that energy density perturbations associated with
the matter sector do decay in all wavelength regimes in contrary to ordinary scalar field models.

As a more ambitious problem which is of great importance is a
construction of the formalism analogous to presented in this paper for
a model with self-interacting non-local scalar field. Such models play
important role in the SFT. For instance, rolling tachyon dynamics is
governed by action (\ref{action_model}) with a polynomial potential of
fourth degree. However, even background solutions are not very well
understood because there is no general analytic way of solving
non-local non-linear equations. On the other hand it follows from the
present analysis that passing to a local system with many fields is
vital for the construction of perturbation equations.

Looking a step further it is interesting to consider perturbations in
other non-local models coming from the SFT. For instance, models where
open and closed string modes are non-minimally coupled may be of
interest in cosmology. An example of the classical solution is
presented in \cite{AKd}. The formalism, presented in this paper, has been
modified~\cite{BKMV_perturbation} to analyse the first order scalar perturbation in the non-local
gravity model, proposed in~\cite{BMS}. We hope that it would be possible to extend
this formalism to other models involving
non-localities like modified gravity setups~\cite{nonlocal}.

The authors are grateful to I.Ya.~Aref'eva, B.~Craps, B.~Dragovich, and
V.F.~Mu\-kha\-nov for useful comments and discussions. This work is
supported in part by Russian Foundation for Basic Research (RFBR) grants
08-01-00798 and 11-01-00894. A.K. is supported in part by the
Belgian Federal Science Policy Office through the Interuniversity
Attraction Poles IAP VI/11, the European Commission FP6 RTN programme
MRTN-CT-2004-005104 and by FWO-Vlaanderen through the project
G.0428.06. Research of S.V. is supported in part by grants of Russian  Ministry of
Education and Science NSh-3920.2012.2,  and in part by contract CPAN10-PD12 (ICE, Barcelona, Spain).


\appendix

\section{Cosmological perturbations formalism}




\subsection{Perturbations in models with several perfect fluids or local scalar fields}

In this Appendix we briefly remind the main equations of the perturbations analysis with  several perfect fluids or local scalar fields
(see details, for example, in paper~\cite{hwangnoh}).

We assume non-interacting scalar fields resulting in individual
conservation equations
\begin{equation}
T=\sum_iT_i,\qquad D_\mu{T_{i}}^{\mu}_{\nu}=0. \label{conservation_i}
\end{equation}
Energy densities and pressures also acquire index $i$ and
\begin{equation}
\varrho=\sum\limits_i\varrho_{i},\qquad p=\sum\limits_ip_{i}.
\end{equation}
 To describe
energy--momentum tensor perturbations we introduce individual quantities
accompanied with index $i$ and the following summation rules hold
\begin{equation}
\begin{split}
\delta\!\varrho&=\sum_i\delta\!\varrho_{i},\qquad \delta\! p=\sum_i\delta\! p_{i},\\
(\varrho+p)v^s&=\sum_i(\varrho_{i}+p_{i})v_{i}^s,\qquad \pi^s=\sum_i\pi^s_{i}.
\end{split}
\label{deltastressi}
\end{equation}
The following additional notations are useful:
\begin{equation*}
\begin{split}
w_{i}&\equiv p_{i}/\varrho_{i},\quad {c_s^2}_{i}\equiv\dot
p_{i}/\dot\varrho_{i},\\e_{i}&\equiv\delta\!
p_{i}-{c_s^2}_{i}\delta\!\varrho_{i},\quad
\delta_{i}\equiv\delta\!\varrho_{i}/\varrho_{i}.
\end{split}
\end{equation*}

For individual fluids one can define the following gauge invariant
quantities:
\begin{equation}
{v_{i}}_\chi=v_{i}^s-\frac
ka\chi,\quad\varepsilon_{i}=\delta_{i}+3(1+w_{i})H\frac akv_{i}^s
\label{GIvarsi}
\end{equation}

Starting with the Einstein equations $G_{\mu\nu}=8\pi G
T_{\mu\nu}$ supplemented with equations
\begin{equation}
\dot\varrho_{i}+3H(\varrho_{i}+p_{i})=0
\end{equation}
for all $i$ one yields an analog of equation (\ref{deltaGIconsa}) for $i$-th fluid
\begin{equation}
\dot {v_{i}}_\chi+H{v_{i}}_\chi=\frac k{a(1+w_{i})}\left(\frac
{e_{i}}{\varrho_{i}}+{c_s^2}_i\varepsilon_{i}+\Phi(1+w_{i})-\frac{2\pi_{i}^{s}}{3\varrho_{i}}\right).
\label{deltaGIconsai}
\end{equation}

We note the only change is that everything that can carry a fluid index
$i$ has acquired it. An analog of equation (\ref{deltaGIcons0}) is not
so straightforward and is given by
\begin{equation}
\begin{split}
&\dot \varepsilon_{i}-3Hw_i\varepsilon_{i}+\frac
ka(1+w_{i}){v_{i}}_\chi\left(1-3\dot
H\frac{a^2}{k^2}\right)+2H\frac{\pi_{i}^{s}}{\varrho_{i}}={}\\&={}-\frac
ka(1+w_{i})3\dot H\frac{a^2}{k^2}v_\chi.
\end{split}
\label{deltaGIcons0i}
\end{equation}
For scalar fields and perfect
fluids anisotropic stresses tensors $\pi_i^s$ are equal to zero. Taking all $\pi_i^s=0$ one has~\cite{addon}:
\begin{equation}
\begin{split}
&\ddot\varepsilon_{i}+\dot\varepsilon_{i}H\left(2+3{c_s^2}_i-6w_{i}\right)+\\
+&\varepsilon_{i}\left(-3\dot H({c_s^2}_i+w_{i})+9H^2{c_s^2}_i-15H^2w_{i}+\frac{k^2}{a^2}{c_s^2}_i\right)=\\
=&-\frac{k^2}{a^2}\frac{e_{i}}{\varrho_{i}}+ \frac{12\pi
G}{\varrho_{i}}\sum_m\left((\varrho_{i}+p_{i})e_{m}-(\varrho_{m}+p_{m})e_{i}\right)
+\\
+&4\pi G(1+w_{i})\sum_m\varrho_{m}\varepsilon_{m}(1+3{c_s^2}_m)+\\
+&\frac{12\pi GH}{3\dot H-\frac{k^2}{a^2}}\sum_m\left[\varrho_{m}(1
+3{c_s^2}_m)\left((1+w_{m})(\dot\varepsilon_{i}-3Hw_{i}\varepsilon_{i})\right.\right.
-\\
&\left.\left.\qquad\quad~-(1+w_{i})(\dot\varepsilon_{m}-3Hw_{m}\varepsilon_{m})\right)\right].
\end{split}
\label{deltaGIepsirecasti0}
\end{equation}

In the case of many free local massive scalar fields we consider action
(\ref{action_model_local}). To the background order one uses
(\ref{EQUFr1}). To the perturbed order one has
\begin{equation}
\begin{split}
\delta\!\varrho_i&=\Fc'(J_i)\left(\dot\tau_i\dot{\delta\!\tau_i}
-\Phi\dot\tau_i^2+J_i\tau_i\delta\!\tau_i\right),\\
\delta\!p_i&=\Fc'(J_i)\left(\dot\tau_i\dot{\delta\!\tau_i}-\Phi\dot\tau_i^2-J_i\tau_i\delta\!\tau_i\right),\\
v_i&=\frac ka\frac{\delta\!\tau_i}{\dot\tau_i},\qquad\pi_i^{s}=0.
\label{trivial}
\end{split}
\end{equation}
It is easy to show that $e_i=(1-{c_s^2}_i)\varrho_i\varepsilon_i $ and
using equation (\ref{deltaGIepsirecasti0}) one gets
\begin{equation}
\begin{split}
&\ddot\varepsilon_{i}+\dot\varepsilon_{i}H\left(2+3{c_s^2}_i-6w_{i}\right)+\\
+&\varepsilon_{i}\left(-3\dot H(1+w_{i})-15H^2w_{i}+9H^2{c_s^2}_i+\frac{k^2}{a^2}\right)=\\
=&16\pi G(1+w_{i})\sum_m\varrho_{m}\varepsilon_{m}+\\
+&\frac{12\pi GH}{3\dot
H-\frac{k^2}{a^2}}\sum_m\left[\varrho_{m}(1+3{c_s^2}_m)
\left((1+w_{m})(\dot\varepsilon_{i}-3Hw_{i}\varepsilon_{i})\right.\right.
-\\
&\left.\left.\qquad\quad~-(1+w_{i})(\dot\varepsilon_{m}-3Hw_{m}\varepsilon_{m})\right)\right].
\end{split}
\label{deltaGIepsirecasti0manyscalar}
\end{equation}
These are the equations governing perturbations of the energy density
if we take as the background solution (\ref{tau_sum}) with arbitrary
number of summands. If we have a mixture of perfect fluids and scalar
fields one can easily compose a system of equations with one subset
representing perturbations of perfect fluids and another subset
representing the perturbations of scalar fields. The cosmological constant can be consider
 a part of a scalar field
potential. Since there was only one scalar field in the original problem we
are mainly interested in the behavior of the perturbation of the total
energy--momentum tensor of scalar fields. Using a relation
\begin{equation}
\varrho\varepsilon=\sum_m\varrho_m\varepsilon_m=\varrho_\tau\varepsilon_\tau
\label{epsrel}
\end{equation}
we see that $\varepsilon_\tau$ as well as total $\varepsilon$ can be
easily extracted. Here by the subscript $\tau$ we denote the total
scalar fields quantities.

Different approach is deriving a system of equations which manifestly
contains one equation for $\varepsilon$. It is possible if one takes as
perturbation variables $\varepsilon$ and
\begin{equation}
\zeta_{ij}\equiv\frac{\delta\!\tau_i}{\dot\tau_i}-\frac{\delta\!\tau_j}{\dot\tau_j}.
\end{equation}
The latter variables are manifestly gauge invariant.

In our case each background
scalar fields satisfy equations (\ref{equtaui}).
Perturbating these equations, one obtains
\begin{equation*}
\ddot{\delta\!\tau}_i+3H\dot{\delta\!\tau}_i+\frac{k^2}{a^2}\delta\!\tau_i
+J_i\delta\!\tau_i=2\alpha\ddot\tau_i+(\kappa+3H\alpha+\dot\alpha)\dot\tau_i
\end{equation*}
where $\kappa=3(-\dot\varphi+H\alpha)+\frac{k^2}{a^2}\chi$. Subtracting
two of such equations for indexes $i$ and $j$ one gets equations (\ref{deltaijeps}).

If $\tau_i=0$ for some $i$,
corresponding $i$-s perturbation variables turn out to be trivial as
well according to (\ref{trivial}). In other words it is not possible at
least at linear order that modes which are trivial in the background
affect perturbations.


\subsection{Space homogeneous perturbations, $k=0$}

Independence of perturbations of spatial
coordinates implies $(0,a)$ and $(a,b)$ for $a\neq b$ components of the
Einstein equation and $a$ components of the conservation equation
remain unperturbed. Effectively this means $\beta=\gamma\equiv0$ in the
metric perturbation (\ref{deltametric}) and $v\equiv0$ in the
energy--momentum tensor perturbation (\ref{deltastresstext}) (the
anisotropic stress $\pi^s=0$ as usual). Therefore,
starting with the Einstein equations $G_{\mu\nu}=8\pi G T_{\mu\nu}$ one
gets only three equations which are as follows
\begin{eqnarray}
\dot{\delta\!\varrho}+3H(\delta\!\varrho+\delta\!p)&=&{}-3\dot\varphi(\varrho+p),\label{k01}\\
-3H(-\dot\varphi+H\Phi)&=&4\pi G\delta\!\varrho,\\
-\ddot\varphi+H\dot\Phi&=&4\pi G(\delta\!\varrho+\delta\!p)
\end{eqnarray}
where the third equation is a consequence of the first two. It is not a
problem to have one equation less since one of the perturbation
functions can be gauged away. In a system with many fluids one has
instead of (\ref{k01})
\begin{eqnarray}
\dot{\delta\!\varrho_i}+3H(\delta\!\varrho_i+\delta\!p_i)+3\dot\varphi(\varrho_i+p_i)&=&0.\label{k01i}
\end{eqnarray}
Note that $\chi$ used before is identically zero and thus cannot be
used to produce the gauge invariant Bardeen potentials. Introducing an
analog of $\varepsilon_i$ which is now
$\bar\varepsilon_i=\delta_i+3\varphi(1+w_i)$ one gets out of the first
equation of the above system
\begin{equation}
\dot{\bar\varepsilon}_i+3H\bar\varepsilon_i({c_s^2}_i-w_i)+3H\frac
{e_i}{\varrho_i}=0.\end{equation} In the latter equation all the
quantities are gauge invariant. All equations are homogeneous and for
any perfect fluid with ${c_s^2}_i=w_i=\const_i$ and $e_i=0$ one gets
$\bar\varepsilon_i=\const$. For instance, for the cosmological constant one has
$\bar\varepsilon_\Lambda=\delta_\Lambda=\const\not\equiv0$. We see that
perturbation of the energy density of the cosmological constant is not obligatory zero
unlike the consideration in the previous Subsections. This happens because
equation (\ref{deltaGIconsai}) used before to claim it is zero is not
applicable for space homogeneous perturbations. Such a situation for
adiabatic perturbations was noted in \cite{Bardeen} with further
reference to \cite{lk}. Scalar field however is an example of a perfect
fluid with entropic perturbations, i.e. $e\neq0$. One can conveniently
introduce the variable
$\varepsilon_i=\delta_i+3H(1+w_i)\frac{\delta\!\tau_i}{\dot\tau_i}$. In
this variables equation (\ref{k01i}) for the perturbation of the energy
density of a scalar field in a system with many scalar fields
(\ref{action_model_local}) becomes
\begin{equation}
\begin{split}
&\frac d{dt}({\varrho_i\varepsilon_i})+3H(\varrho_i\varepsilon_i)+\\
+&(\varrho_i+p_i)\left(\frac{4\pi
G}H\left(\sum_m\varrho_m\varepsilon_m+\sum_j\delta\!\varrho_j\right)-{}\right.\\
 {}-&\left. 3\dot H\frac{\delta\!\tau_i}{\dot\tau_i}-12\pi
G\sum_m\Fc'(J_k)\dot\tau_k\delta\!\tau_k\right)=0.
\end{split}
\label{k0is}
\end{equation}
Here in the second row
$\sum_m\varrho_m\varepsilon_m=\varrho_\tau\varepsilon_\tau$ is the
summation over all scalar fields components and
$\sum_j\delta\!\varrho_j$ is the summation over all other fluids. In
the most interesting and important case when apart from scalar fields
there is only the cosmological constant this summation over $j$ is just a single constant
term. Moreover, since for the cosmological constant $\varepsilon=\bar\varepsilon$ one can
write this constant term as $\varrho_\Lambda\varepsilon_\Lambda$ (here
$\rho_\Lambda=g_o^2\Lambda$). Further, defining $\varepsilon$ through
$\varrho\varepsilon=\varrho_\tau\varepsilon_\tau+\varrho_\Lambda\varepsilon_\Lambda$
and summing up all the equations (\ref{k0is}) for the scalar fields one
derives
\begin{equation*}
\dot\varepsilon+3H\varepsilon\left(1+\frac{\dot
H}{3H^2}\right)=\frac{8\pi G\varrho_\Lambda\varepsilon_\Lambda}{H}.
\end{equation*}
The latter equation can be integrated in terms of $a$ and $H$ to give
\begin{equation}
{\varepsilon}=\frac1{Ha^3}\left(\varepsilon_0+8\pi G\varrho_\Lambda\varepsilon_\Lambda\int a^3dt\right).
\end{equation}
The quantity of interest $\varepsilon_\tau$ can be restored using
$\varepsilon_\tau=(\varrho\varepsilon-\varrho_\Lambda\varepsilon_\Lambda)/\varrho_\tau$.

\end{document}